\documentclass[
    prb,
    twoside,
    twocolumn,
    10pt,
    floatfix,
    showpacs,
    citeautoscript,
]{revtex4-1}

\usepackage{amsmath}
\usepackage{amssymb}
\usepackage{glossaries}
\usepackage{graphicx}
\usepackage{times}
\usepackage{units}
\usepackage[dvipsnames]{xcolor}

\usepackage[colorlinks=true]{hyperref}
\hypersetup{
    colorlinks=true,
    linkcolor=NavyBlue,
    citecolor=NavyBlue,
    filecolor=NavyBlue,
    urlcolor=NavyBlue,
}

\newcommand{\dynasor}{\textsc{dynasor}}
\renewcommand{\vec}[1]{\ensuremath\mathbf{#1}}

\newacronym{bcc}{BCC}{body-centered cubic}
\newacronym{dft}{DFT}{density-functional theory}
\newacronym{dos}{DOS}{density of states}
\newacronym{eam}{EAM}{embedded-atom method}
\newacronym{fc}{FC}{force constant}
\newacronym{fcc}{FCC}{face-centered cubic}
\newacronym{hcp}{HCP}{hexagonal-closed packed}
\newacronym{la}{LA}{longitudinal acoustic}
\newacronym{md}{MD}{molecular dynamics}
\newacronym{sed}{SED}{spectral energy density}
\newacronym{ta}{TA}{transverse acoustic}
\newacronym{vacf}{VACF}{velocity autocorrelation function}

\begin{document}

\title{
    \dynasor{} -- A tool for extracting dynamical structure factors \texorpdfstring{\\}{}
    and current correlation functions from molecular dynamics simulations
}

\author{Erik Fransson}
\author{Mattias Slabanja}
\author{Paul Erhart}
\email{erhart@chalmers.se}
\author{G\"oran Wahnstr\"om}
\email{goran.wahnstrom@chalmers.se}
\affiliation{
  Chalmers University of Technology,
  Department of Physics,
  S-412 96 Gothenburg, Sweden
}

\begin{abstract}
Perturbative treatments of the lattice dynamics are widely successful for many crystalline materials, their applicability is, however, limited for strongly anharmonic systems, metastable crystal structures and liquids.
The full dynamics of these systems can, however, be accessed via molecular dynamics (MD) simulations using correlation functions, which includes dynamical structure factors providing a direct bridge to experiment.
To simplify the analysis of correlation functions, here the \dynasor{} package is presented as a flexible and efficient tool that enables the calculation of static and dynamical structure factors, current correlation functions as well as their partial counterparts from MD trajectories.
The \dynasor{} code can handle input from several major open source MD packages and thanks to its C/Python structure can be readily extended to support additional codes.
The utility of \dynasor{} is demonstrated via examples for both solid and liquid single and multi-component systems.
In particular, the possibility to extract the full temperature dependence of phonon frequencies and lifetimes is emphasized.
\end{abstract}

\maketitle

\section{Introduction}

The dynamical properties of materials are fundamental to, e.g., their thermodynamic, kinetic, optical and transport properties.
They can be accessed via neutron \cite{PetHeiTra91, ChrAbrChr08, LinErhBet18, LiHellMa2014} or X-ray \cite{Bar20} scattering experiments, which provide quantitative information in the form of dynamical structure factors \cite{Lov84, HanMcD06}.
The latter can also be generated using atomic scale modeling via \gls{md} simulations or lattice dynamics, providing a quantitative bridge between experiment and atomic scale modeling.

\Gls{md} simulations are the primary choice for modeling liquids \cite{AllTil87} and in recent years several packages geared toward the analysis of their dynamics have emerged, including, e.g., \textsc{nMoldyn} \cite{RogMurHin2003}, \textsc{mdanse} \cite{GorAouPel2017}, \textsc{liquidlib} \cite{WalJaiCai2018}, and \textsc{freud} \cite{RamDicHar2020}.
The dynamical properties of (periodic) solid state systems are, on the other hand, commonly analyzed within the framework of lattice dynamics, i.e. a low-order expansion of energy and forces in terms of small atomic displacements.
The lowest (second-order) \gls{fc} expansion can be conveniently handled using packages such as \textsc{phonopy} \cite{TogTan15} or \textsc{phonon} \cite{ParLiKaw97}, while higher-order terms can be obtained either directly via tools such as \textsc{phono3py} \cite{TogChaTan15} and \textsc{shengbte} \cite{LiCarKat14} (for third-order terms) or via regression using, e.g., \textsc{alamode} \cite{TadGohTsu14}, \textsc{tdep} \cite{HelAbrSim11}, \textsc{csld} \cite{ZhoNieXia14} or \textsc{hiphive} \cite{EriFraErh19}.

The calculation of the dynamical properties in general and the dynamical structure factor in particular via the \gls{fc} approach typically includes only second-order (harmonic) or third-order (lowest anharmonic order) terms, limiting the approach to materials with relatively weak anharmonicity.
Moreover, the quick explosion of terms with system size imposes a rather severe limit on system size.
As a result, the computation of the dynamical properties becomes very cumbersome or impossible for materials with large unit cells, low symmetry and/or strong anharmonicity, including metastable crystal structures and materials with soft modes, which exhibit particular rich and interesting dynamical properties.
\footnote{
    For lattice-based systems phonon frequencies and lifetimes can also be obtained from \gls{md} simulations via the spectral energy density \cite{ThoTurIut10, DinPeiJia15}.
    The latter approach, however, breaks down for amorphous structures and solids that exhibit diffusion.
}

All of the latter limitations can in principle be overcome by analyzing correlation functions, such as the dynamical structure factor, from \gls{md} simulations using forces from \gls{dft} calculations, empirical potentials \cite{Pli95} or high-order force constants \cite{FraEriErh20}.
To take full advantage of this approach it is desirable to obtain the dispersion relations as a function of not only the magnitude but also the direction of the momentum transfer vector.
While this information is in principle present whenever analyzing trajectories from periodic systems, this is not the primary focus of the aforementioned tools.
Here, to fill this need, we introduce the \dynasor{} package for the efficient calculation of dynamical structure factors from \gls{md} trajectories.
While it is generally applicable to both solids and liquids, it is particularly well suited to analyze the dynamics of fully or partially periodic systems.

\dynasor{}, which is written in a combination of C and Python, can parse \gls{md} trajectories from \textsc{lammps} \cite{Pli95}, \textsc{gromacs} \cite{AbrMurSch15} as well as \textsc{namd} \cite{PhiBraWan05}, and can be extended straightforwardly to support additional formats.
If \textsc{vmd} \cite{vmd} is available, \dynasor{} can employ the \texttt{molfileplugin} to read even more formats (with some limitations).
The code then allows one to compute not only the dynamical structure factor but also current correlation as well as partial correlation functions.
In this fashion, it is for example possible to extract the full temperature dependence of phonon dispersions, as illustrated below for both solids (\autoref{sect:solid-aluminum} and \ref{sect:bcc-Ti}) and liquids (\autoref{sect:liquid-aluminum} and \ref{sect:sodium-chloride}).

Below, we first provide a review of the theoretical background, before describing the implementation and basic usage of \dynasor{}.
We then demonstrate the application and performance of the code for both solids and liquids, and specifically illustrate the extraction of phonon dispersions and lifetimes.

\section{Theoretical background}

In the following we provide a concise compilation of the expressions for the dynamical structure factor and current correlation functions in terms of the atomic coordinates and velocities.
More extensive information can be found, e.g., in Refs.~\cite{HanMcD06} and \cite{BooYip80}.
We describe a theoretical framework for how these correlation functions can be analyzed in order to extract vibrational information of the system.

\subsection{Dynamical structure factor}
The density of atoms $n(\vec{r},t)$ is defined as
\begin{align*}
    n(\vec{r},t) = \sum_i^N \delta (\vec{r} - \vec{r}_i(t)),
\end{align*}
where $\vec{r}_i(t)$ denotes the position of atom $i$ at time $t$ and $N$ is the total number of atoms.
The density can be spatially Fourier transformed via
\begin{align}
    n(\vec{q}, t) = \int n(\vec{r}, t)\mathrm{e}^{\mathrm{i}\vec{q} \cdot \vec{r}} \mathrm{d}\vec{r} = \sum _i ^N \mathrm{e}^{\mathrm{i}\vec{q} \cdot \vec{r}_i(t)}.
\label{eq:particle_density}
\end{align}
The intermediate scattering function $F(\vec{q},t)$ is defined in terms of the time correlation function of $n(\vec{q}, t)$ as
\begin{align*}
    F(\vec{q},t) = \frac{1}{N}\left<n(\vec{q},t)n(-\vec{q},0)\right>,
\end{align*}
where $\left <\ldots\right >$ denotes an ensemble average or ---if the systems is ergodic--- a time average.
The static structure factor is given by the initial value of the intermediate scattering function
\begin{align*}
    S(\vec{q})=F(\vec{q},t=0),
\end{align*}
while one obtains the dynamical structure factor $S(\vec{q},\omega)$ via a temporal Fourier transformation of $F(\vec{q},t)$
\begin{align*}
    S(\vec{q},\omega) = \int_{-\infty}^\infty F(\vec{q},t)\mathrm{e}^{-\text{i}\omega t} \mathrm{dt}.
    \label{eq:Skw}
\end{align*}
$S(\vec{q},\omega)$ exhibits peaks in the $(\vec{q},\omega)$ plane corresponding to longitudinal modes.
The broadening of these peaks is related to the phonon lifetimes and thus the anharmonicity of the system (\autoref{sect:solid-aluminum} below).

\subsection{Velocity autocorrelation function}
The \gls{vacf}, $\Phi(t)$, is defined as
\begin{align*}
    \Phi(t) = \frac{1}{N}\sum_i ^N\frac{\left<\vec{v}_i(t)\cdot\vec{v}_i(0)\right>}{\left<\vec{v}_i(0)\cdot \vec{v}_i(0)\right>},
\end{align*}
where ${v}_i(t)$ denotes the velocity of atom $i$ at time $t$.
The Fourier transformation of $\Phi(t)$ is related to the vibrational density of states, $g(\omega)$, via
\begin{equation}
g(\omega) = \frac{2}{\pi}\int_{0}^\infty \Phi(t)\cos(\omega t) \mathrm{dt}.
\label{eq:dos_from_vacf}
\end{equation}

\subsection{Current correlations}
In order to obtain mode specific vibrational frequencies the positions of the atoms need to be included in the analysis.
This can be done by computing current correlation functions.
These are defined in a fashion that is analogous to the approach for the intermediate scattering function, but with the atom density being replaced with the current density, $n(\vec{r},t)$,
\begin{align*}
    &\vec{j}(\vec{r},t) = \sum_i^N \vec{v}_i(t) \delta (\vec{r} - \vec{r}_i(t)) \\
    &\vec{j}(\vec{q},t) = \sum_i^N \vec{v}_i(t) \mathrm{e}^{\mathrm{i}\vec{q} \cdot \vec{r}_i(t)}.
\end{align*}
The current density is a vector quantity which can be decomposed into a longitudinal part containing the component parallel to the $\vec{q}$-vector and a transverse part containing the perpendicular component, according to
\begin{align}
    \vec{j}(\vec{q},t) = \vec{j}_L(\vec{q},t) + \vec{j}_T(\vec{q},t),
    \label{eq:current_density}
\end{align}
where
\begin{align*}
    &\vec{j}_L(\vec{q},t) = \sum_i ^N(\vec{v_i}(t) \cdot\hat{\vec{q}})\hat{\vec{q}}\mathrm{e}^{\mathrm{i}\vec{q} \cdot \vec{r}_i(t)} \\
    &\vec{j}_T(\vec{q},t) = \sum_i ^N \left[\vec{v_i}(t) - (\vec{v_i}(t) \cdot \hat{\vec{q}}) \hat{\vec{q}}\right] \mathrm{e}^{\mathrm{i}\vec{q} \cdot \vec{r}_i(t)}
\end{align*}
and $\hat{\vec{q}}$ denotes the unit vector.
The current correlation functions can now be computed (analogous to the intermediate scattering function) as
\begin{align*}
    &C_L(\vec{q},t) = \frac{1}{N}\left<\vec{j}_L(\vec{q},t)\cdot\vec{j}_L(-\vec{q},0)\right> \\
    &C_T(\vec{q},t) = \frac{1}{N}\left<\vec{j}_T(\vec{q},t)\cdot\vec{j}_T(-\vec{q},0)\right>.
\end{align*}
As in the case of the intermediate scattering function, the current correlations can be temporally Fourier transformed to the frequency domain.
By inspection of \eqref{eq:particle_density} and \eqref{eq:current_density} particle density and current density are related via
\begin{align*}
    \frac{\partial }{\partial t}n(\vec{q}, t) = \mathrm{i} \vec{q}\cdot \vec{j}(\vec{q}, t),
\end{align*}
which yields the following relation
\begin{align}
    \omega^2 S(\vec{q},\omega) = q^2 C_L(\vec{q},\omega)
    \label{eq:Jl_S_relation}
\end{align}
in the frequency domain.

\subsection{Multi-component systems and liquids}
In multi-component systems one can furthermore introduce partial correlation functions.
For example in the case of a binary system (AB) the above expressions for the particle density generalize to
\begin{align*}
    &n_\text{A}(\vec{q}, t) = \sum_i ^{N_\text{A}} \mathrm{e}^{\mathrm{i}\vec{q} \cdot \vec{r}_i(t)} \\
    &F_\text{AB}(\vec{q}, t) = \frac{1}{\sqrt{N_\text{A}N_\text{B}}} \left<n_\text{A}(\vec{q},t)n_\text{B}(-\vec{q},0)\right>.
\end{align*}
This generalization extends to current correlations in the same manner.
In some situations, instead of analyzing the partial correlation functions directly, it is convenient to consider linear combinations of these functions.
This will be demonstrated and discussed in the case of liquid NaCl in \autoref{sect:sodium-chloride}.

In solids, it is often desirable to determine the above mentioned quantities along specific paths connecting high symmetry $\vec{q}$-points.
In isotropic samples on the other hand, such as for example liquids, it is usually preferable to compute these functions with respect to $q=|\vec{q}|$, by performing a spherical average.

\subsection{Damped harmonic oscillators -- Fitting}
\label{sect:fitting_procedure}

Phonons are often modeled as damped harmonic os\-cil\-la\-tors \cite{MeyEnt98, PetHeiTra91}.
This enables correlation functions from both experiments and computer simulations to be fitted to the corresponding analytic functions, allowing the extraction of phonon frequency and lifetime (or damping factor).
For this purpose, the analytic form for the above mentioned correlation functions is derived and analyzed in the following.

Assuming the particle density, $n(\vec{q}, t)$, oscillates as a damped harmonic oscillator then $F(\vec{q}, t)$ is, for each $\vec{q}$, described by a function $a(t)$ that is given by
\begin{align*}
    \frac{\mathrm{d}^2 }{\mathrm{d} t^2}a(t) + \Gamma \frac{\mathrm{d} }{\mathrm{d} t}a(t) + \omega_0^2   a(t) = 0,
\end{align*}
where $\Gamma$ is the damping coefficient and $\omega_0$ the natural frequency of the oscillator.
This means that $\Gamma=\Gamma(\vec{q})$ and $\omega_0=\omega_0(\vec{q})$, but for simplicity these arguments are left out through out the rest of this section.
This equation is solved under the assumptions that $\frac{\mathrm{d}}{\mathrm{d} t} a(t=0) = 0$ and $t\geq 0$.
For simplicity we set $a(t=0)=A$ yielding the following solution
\begin{align*}
    \begin{matrix}
        F(t) = A \mathrm{e}^{-\Gamma t/2} \big(\cos{\omega_e t} + \frac{\Gamma}{2\omega_e}\sin{\omega_e t}\big)\,\, , \,\, \omega_0 > \frac{\Gamma}{2}\\
        F(t) = A \mathrm{e}^{-\Gamma t/2} \big(\cosh{\omega_e t} + \frac{\Gamma}{2\omega_e}\sinh{\omega_e t}\big)\,\, , \,\,\omega_0 < \frac{\Gamma}{2}
    \end{matrix},
\end{align*}
where $\omega_e = \sqrt{\omega_0^2 - \frac{\Gamma^2}{4}}$ and $\omega_0 > \frac{\Gamma}{2}$ represents the underdamped limit.
This yields three fitting parameters $A$, $\Gamma$, and $\omega_0$ for each $\vec{q}$.
The functional form for $a(t)$ can be Fourier transformed to
\begin{align*}
    a(\omega) = A\frac{2\Gamma \omega_0^2}{(\omega^2 - \omega_0^2)^2 + (\Gamma\omega)^2}.
\end{align*}
This corresponds to the analytic functional form of the dynamical structure factor, which is thus a peaked function with a maximum at $\omega_\text{max}=\sqrt{\omega_0^2 - \frac{\Gamma^2}{2}}$ and full-width-at-half-maximum $\text{FWHM} \approx \Gamma$.

This analysis can be extended to current correlation functions by considering \eqref{eq:Jl_S_relation}, giving the following solutions
\begin{align*}
    b(\omega) = B\frac{2\Gamma \omega^2}{(\omega^2 - \omega_0^2)^2 + (\Gamma\omega)^2}.
\end{align*}
This is a peaked function with a maximum at $\omega_\text{max}=\omega_0$ and full-width-at-half-maximum $\text{FWHM} \approx \Gamma$.
In the time domain this function becomes
\begin{align*}
   \begin{matrix}
        b(t) = B \mathrm{e}^{-\Gamma t/2} \big(\cos{\omega_e t} - \frac{\Gamma}{2\omega_e}\sin{\omega_e t} \big) , \,\, \omega_0 > \frac{\Gamma}{2}\\
        b(t) = B \mathrm{e}^{-\Gamma t/2} \big(\cosh{\omega_e t} - \frac{\Gamma}{2\omega_e}\sinh{\omega_e t} \big)\,\, , \,\,\omega_0 < \frac{\Gamma}{2}
    \end{matrix},
\end{align*}
with three fit parameters $B$, $\Gamma$, and $\omega_0$.
While these expressions are strictly valid for the longitudinal current correlations, we assume the same functional form also when fitting the transverse components.
Since there are two transverse modes a sum of two functions is needed (unless the transverse mode is degenerate), giving us six fit parameters instead of three.
The damping coefficient $\Gamma$ is related to the phonon lifetime (also referred to as relaxation or scattering time) $\tau$ as $\tau = 2 / \Gamma$.

\subsection{Fourier transforms}
It is often desirable to transform time dependent correlation functions to the frequency domain.
There are many different methods for carrying out numerical Fourier transforms, window functions can be applied and the signal can be zero padded to obtain better accuracy.
Since all time-dependent functions are included in the output from \dynasor{} it is therefore possible to carry out the Fourier transform in any which way.
By default \dynasor{} will provide correlation functions also in the frequency domain using Filon's formula to carry out the transform as described in appendix D of Ref.~\cite{AllTil87}.
We note that using window functions such as a Fermi-Dirac function
\begin{equation}
    h(t) = \frac{1}{\text{e}^{(t-t_0)/t_\text{width}} +1}
\end{equation}
works very well for preserving the important features but reduces the noise in the correlation functions.
Here, $t_0$ and $t_{width}$ are parameters that should be suitably chosen, given the relaxation time of the correlation function.

\section{Software details}

\dynasor{} is distributed under on open source software license (MIT) and its development is hosted on \textsc{gitlab} \cite{gitlab_dynasor}.
A comprehensive documentation written in \textsc{sphinx} \cite{sphinx} is included in the distribution and is also available online \cite{dynasor_website}.
Below some implementation aspects and inner workings of \dynasor{} are outlined.

\begin{figure}
  \centering
  \includegraphics[scale=0.4]{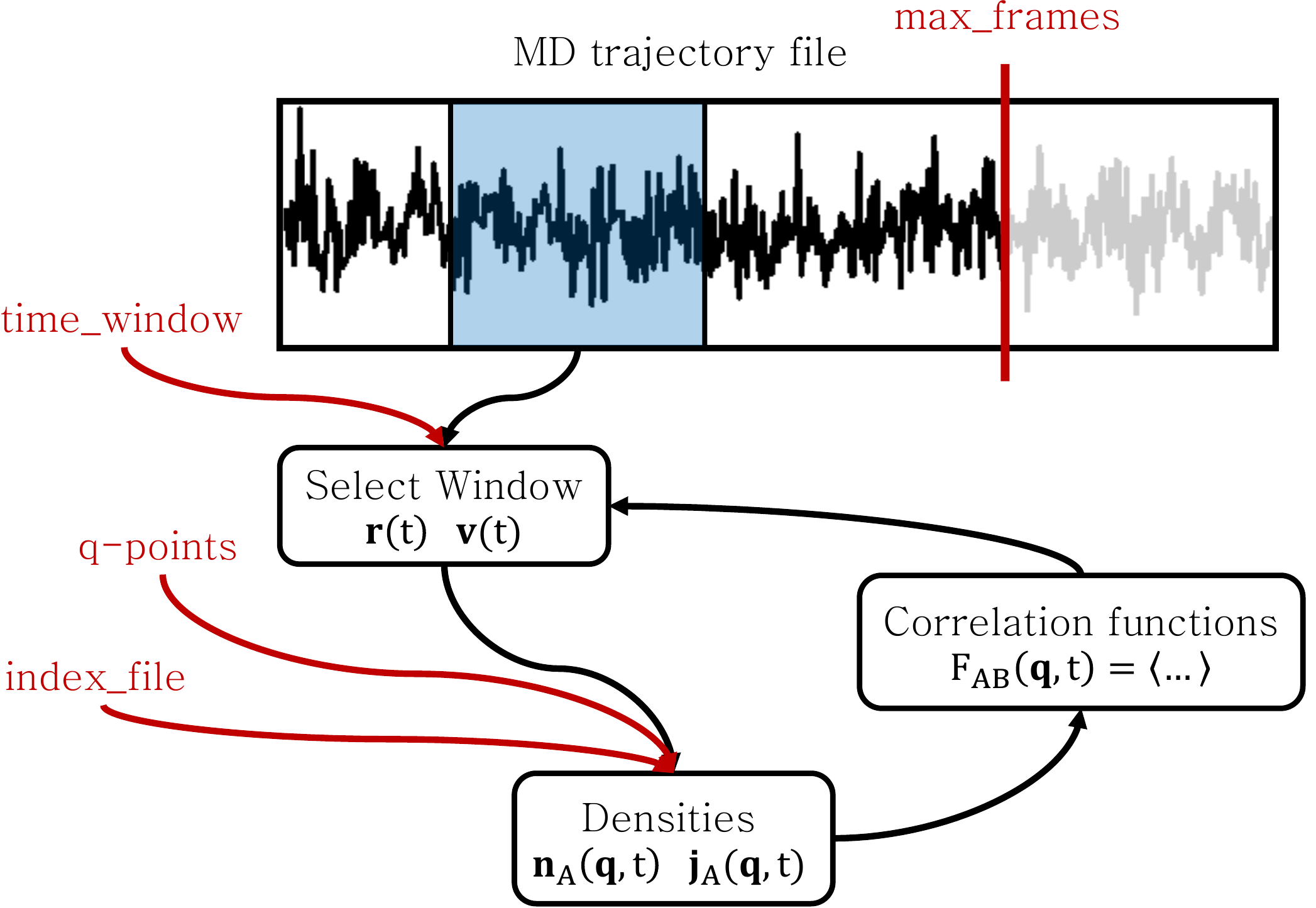}
  \caption{
    Internal workflow of \dynasor{}.
    Text in red marks user inputs.
    The time window, indicated in blue, is moved through the trajectory until the specified maximum number of frames is reached.
  }
  \label{fig:workflow}
\end{figure}

The workflow of \dynasor{} is illustrated in \textbf{\autoref{fig:workflow}}.
The collection of snapshots corresponding to a time window is parsed from the \gls{md} trajectory, and for each snapshot the densities are computed and then the correlation functions.
This process is repeated until there are no more snapshots left in the trajectory or the limit of numbers of snapshots to consider is met.

\dynasor{} can read and parse trajectories in \textsc{lammps} dump format.
If the \textsc{libgmx} library from the \textsc{gromacs} package is available, \dynasor{} can also read \textsc{gromacs} \texttt{xtc}-files.
If \textsc{vmd} is installed \cite{vmd}, \dynasor{} can use the \textsc{molfile} plugin to read other formats (with some limitations) as well.

The time sampling can be adjusted via input parameters such as how \verb|NT| (size of time window) and \verb|MAX_FRAMES| (maximum number of snapshots to consider).
$\vec{q}$-point sampling is configured via four different parameters, controlling sampling style (isotropic or along a path), the maximum $\vec{q}$-vector to include, and the number of $\vec{q}$-points/$\vec{q}$-bins.
For multi-component systems an index file (\verb|index_file|) must be provided, indicating which atomic indices corresponds to which atom types.
For more details about these input parameters see the \dynasor{} documentation or examples \cite{dynasor_website}.

The output data from \dynasor{} consists of
\begin{itemize}
\item Partial intermediate scattering function $F(\vec{q},t)$
\item Partial dynamical structure factor $S(\vec{q},\omega)$
\item Partial longitudinal and transverse partial current correlations $C(\vec{q},t)$ and $C(\vec{q},\omega)$
\item Partial van Hove function $G(\vec{r},t)$
\item Partial self part of $F(\vec{q},t)$ and $S(\vec{q},\omega)$
\end{itemize}
This collection of data can be written as Python pickle-files or \textsc{matlab}/\textsc{octave} \texttt{.m} files.

The computationally most demanding task in the process pipeline, concerns the calculation of the Fourier transformed densities, $n(\vec{q},t)$ and $\vec{j}(\vec{q},t)$.
This part is implemented in C and is accelerated by parallelization using \textsc{OpenMP} \cite{openmp} or \textsc{OpenACC} \cite{acc}.
Once the densities have been computed the averaging of the time correlations is performed in Python.

\section{Applications}
We now turn to exemplary applications of \dynasor{} to ``real'' materials and illustrate the information available in the correlation functions for different systems.
In all cases described below \gls{md} simulations were carried out using the \textsc{lammps} package \cite{Pli95}.
Following equilibration in either the canonical ($NVT$) or isothermal-isobaric ($NPT$) ensemble using the Nos\'e-Hoover thermostat and/or barostats, positions and velocities were sampled for about one nanosecond in the microcanonical ($NVE$) ensemble in order to avoid the thermostat/barostat influencing the dynamics and thus the correlation functions.

When showing dispersion relations, we chose to plot $\omega_0$, rather than $\omega_e$, unless explicitly noted.
The difference between $\omega_0$ and $\omega_e$ is often small, but for system with strong damping there is a clear difference as shall be discussed in the case of \gls{bcc}-Ti (\autoref{sect:bcc-Ti}).

\subsection{Solid (FCC) aluminum}
\label{sect:solid-aluminum}

We first consider \gls{fcc}-Al since it is a rather harmonic system, for which we can carry out meaningful comparisons with perturbative methods.

The atomic interactions in aluminum were modeled using an embedded atom method potential for \cite{MisFarMeh99} and simulations were carried out at \unit[300]{K} and \unit[900]{K} using a supercell comprising $12\times12\times12$ conventional face-centered cubic unit cells.

\begin{figure}
    \centering
    \includegraphics[scale=0.95]{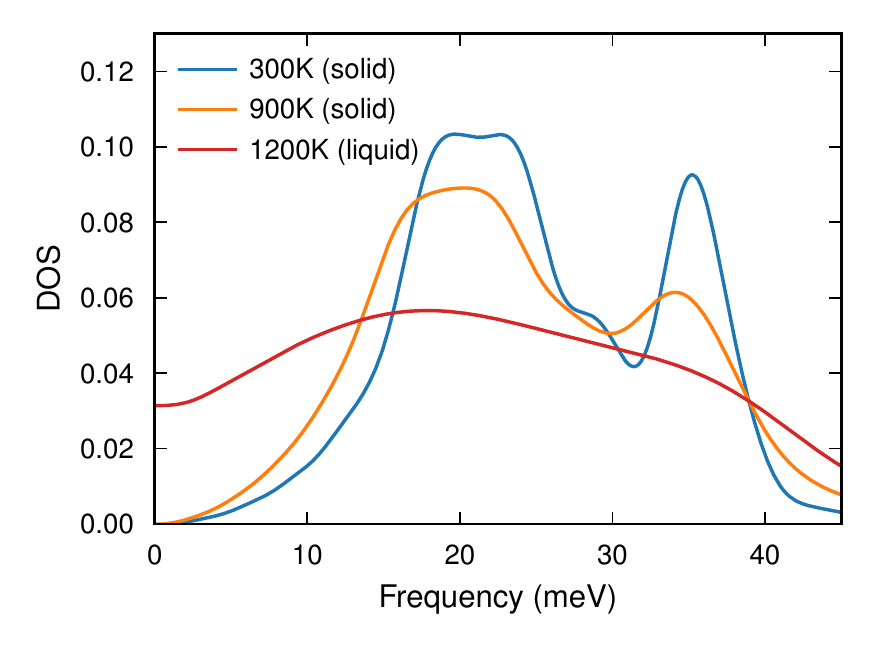}
    \caption{
        Vibrational density of states for aluminum at \unit[300]{K} (solid), \unit[900]{K} (solid), and \unit[1200]{K} (liquid), computed from the velocity autocorrelation function.
    }
    \label{fig:Al_dos}
\end{figure}

A comparison of the \gls{dos} for \gls{fcc} and liquid Al (to be described in more detail in \autoref{sect:liquid-aluminum}), computed via Eq.~\eqref{eq:dos_from_vacf}, shows obvious qualitative differences between the solid and liquid phases (\textbf{\autoref{fig:Al_dos}}).
The \gls{dos} in the solid phase vanishes at zero frequency and shows softening with increasing temperature.
By contrast the liquid \gls{dos} is finite at zero frequency corresponding to diffusive motion while still exhibiting structure at nonzero frequencies resembling solid behavior.

\begin{figure}
    \centering
    \includegraphics[scale=0.95]{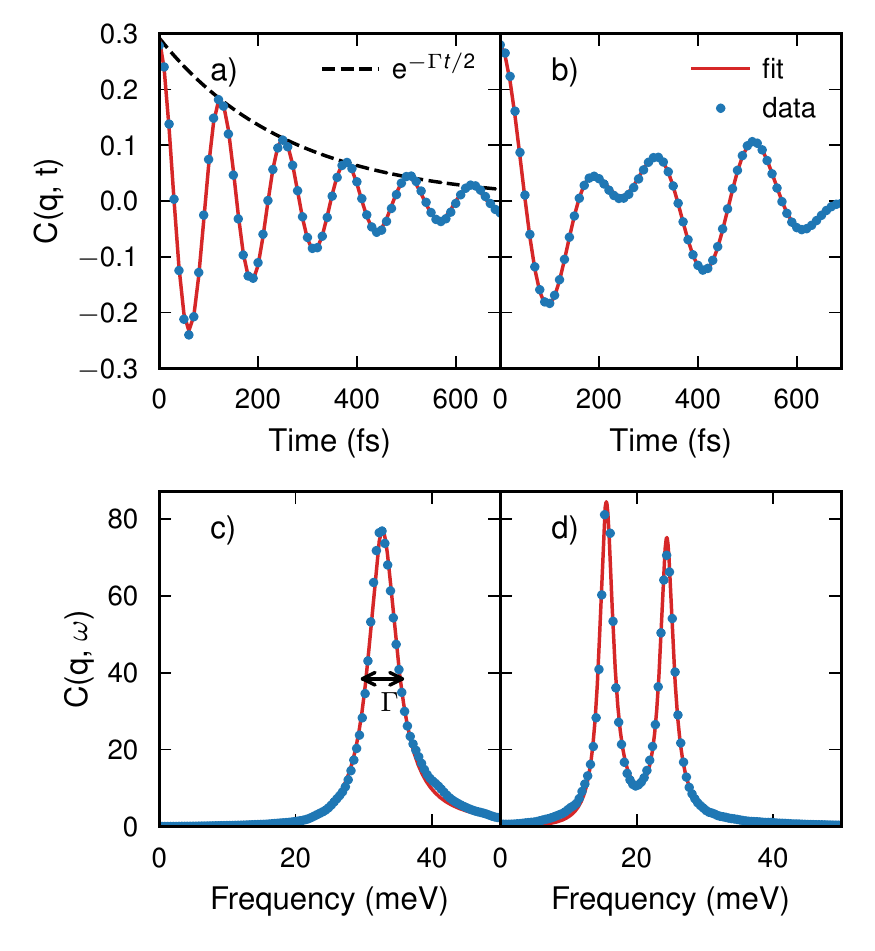}
    \caption{
        Solid (FCC) aluminum at \unit[900]{K}.
        Fits for the (a) longitudinal and (b) transverse current correlations in the time domain for a $\vec{q}$-point halfway along the $\Gamma - K - X$ path.
        The corresponding functions in the frequency domain are shown for the longitudinal (c) and transverse current correlation functions (d).
    }
    \label{fig:Al-fit}
\end{figure}

The phonon frequencies and lifetimes were extracted at a fixed lattice parameter, $a=\unit[4.05]{\text{\AA}}$, across the entire temperature range in order to enable comparison with results from the harmonic approximation and third-order perturbation theory.
The current correlation functions were calculated and fitted in the time domain using the procedure outlined in \autoref{sect:fitting_procedure} for both the longitudinal and transverse current correlation functions (\autoref{fig:Al-fit}).
The representation via the analytical functions in the frequency domain also matches the Fourier transformed data very well, providing further validation of the approach.
While fitting in the time domain is often easier when dealing with few modes, the frequency domain becomes preferable when many modes are present since they are more clearly separated along the $\omega$ axis.

\begin{figure}
    \centering
    \includegraphics[scale=0.95]{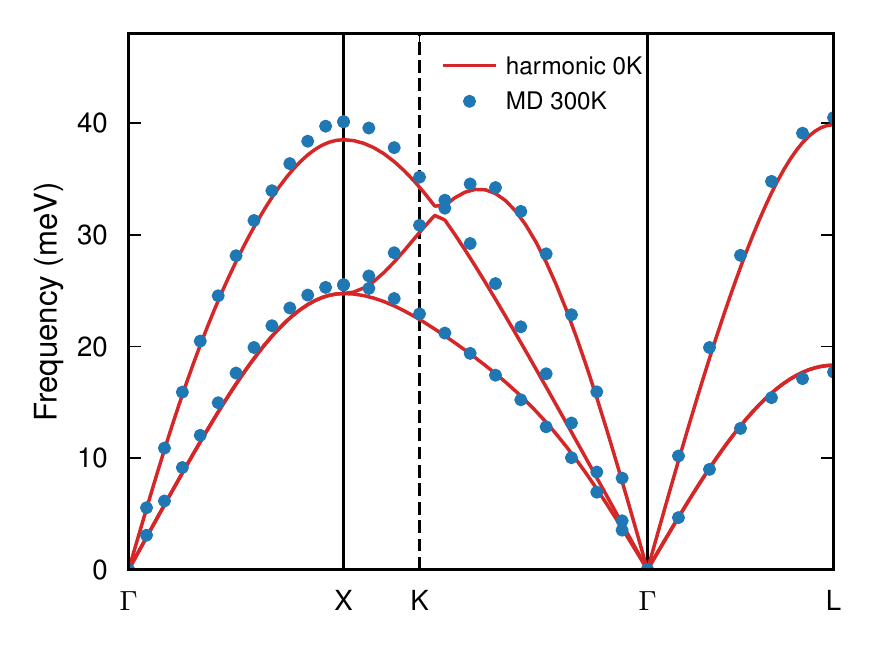}
    \caption{
        Phonon dispersion of solid (FCC) aluminum from \gls{md} simulations at \unit[300]{K} and in the harmonic (\unit[0]{K}) approximation.
    }
    \label{fig:Al_dispersion}
\end{figure}

By extending the fitting procedure to all $\vec{q}$-points in the supercell, one obtains the full phonon dispersion (\autoref{fig:Al_dispersion}).
In the present case, the phonon dispersion at \unit[300]{K} closely agrees with the harmonic (zero Kelvin) phonon dispersion obtained via \textsc{phonopy} \cite{TogTan15}, as expected given the weak anharmonicity in \gls{fcc}-Al.

\begin{figure}
    \centering
    \includegraphics[scale=0.95]{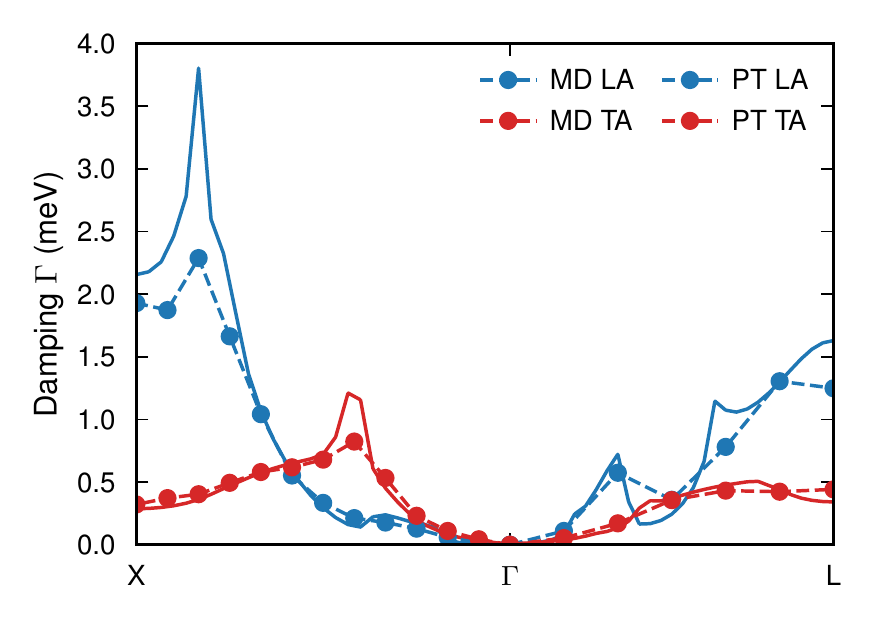}
    \caption{
        Phonon damping (inverse lifetime) in solid (FCC) aluminum at \unit[300]{K} from \gls{md} simulations (via \dynasor{}) and third-order perturbation theory (PT) via \textsc{phono3py}.
    }
    \label{fig:Al_lifetimes}
\end{figure}

The phonon lifetimes calculated using \dynasor{} and \textsc{phono3py} are shown along $\Gamma - X$ and $\Gamma - L$ at \unit[300]{K} in \autoref{fig:Al_lifetimes}. Here, $\Gamma$ obtained from \dynasor{} is shown together with the damping obtained from \textsc{phono3py}.
The latter has been multiplied by a factor of four to accommodate the different definitions.
The phonon lifetime is given by
\begin{equation}
    \tau = \frac{2}{\Gamma_\text{dynasor}} = \frac{1}{2\Gamma_\text{phono3py}}
\end{equation}
and hence for consistency we compare $\Gamma_\text{dynasor}$ with $4\Gamma_\text{phono3py}$.

\begin{figure}
    \centering
    \includegraphics[scale=0.95]{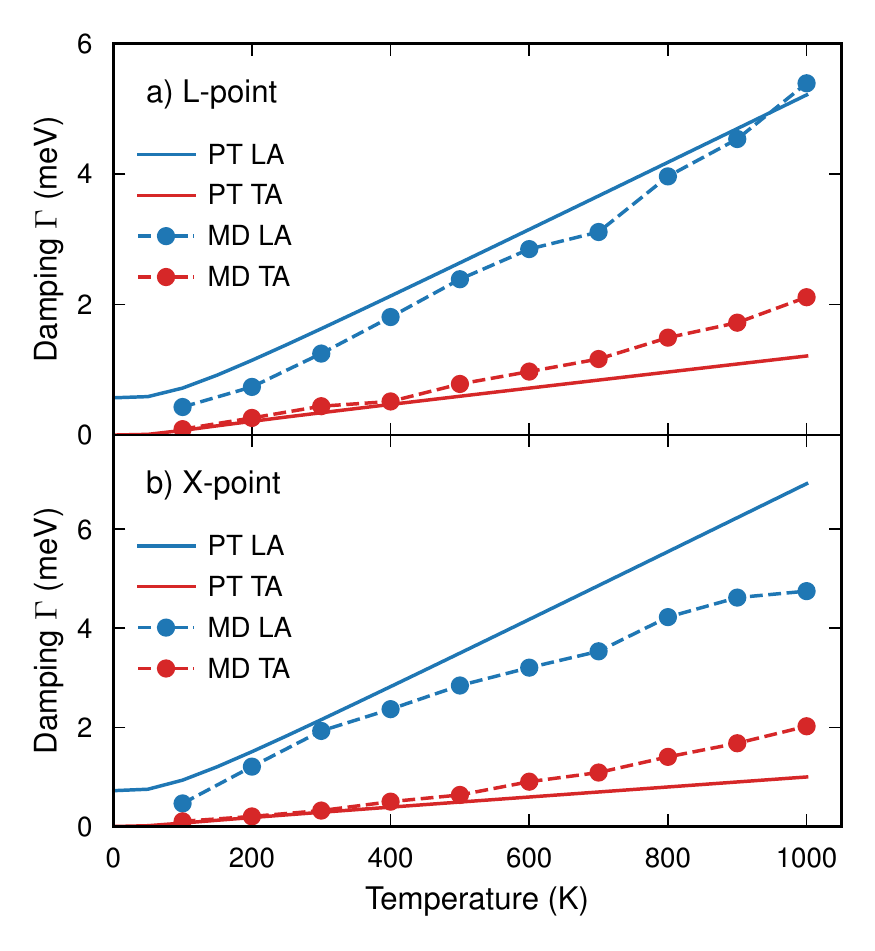}
    \caption{
        Phonon damping (inverse lifetime) in solid aluminum from \gls{md} simulations (via \dynasor{}) and third-order perturbation theory (PT), via \textsc{phono3py}, for the  for the LA and TA modes at (a) L and (b) X as a function of temperature.
    }
    \label{fig:Al_lifetimes_temperature}
\end{figure}

The lifetimes are long, consistent with weak anharmonicity.
The agreement between \dynasor{} and \textsc{phono3py} is good, both qualitatively and quantitatively.
This is expected for low temperatures where third-order perturbation theory captures most of the relevant anharmonic contributions to the lifetimes.
As temperature increases higher-order terms become increasingly important and the lifetimes obtained third-order perturbation theory deviate more and more strongly from those obtained from \gls{md} simulations, which include scattering to all orders (\autoref{fig:Al_lifetimes_temperature}).

\subsection{Liquid aluminum}
\label{sect:liquid-aluminum}

\begin{figure*}
  \centering
  \includegraphics{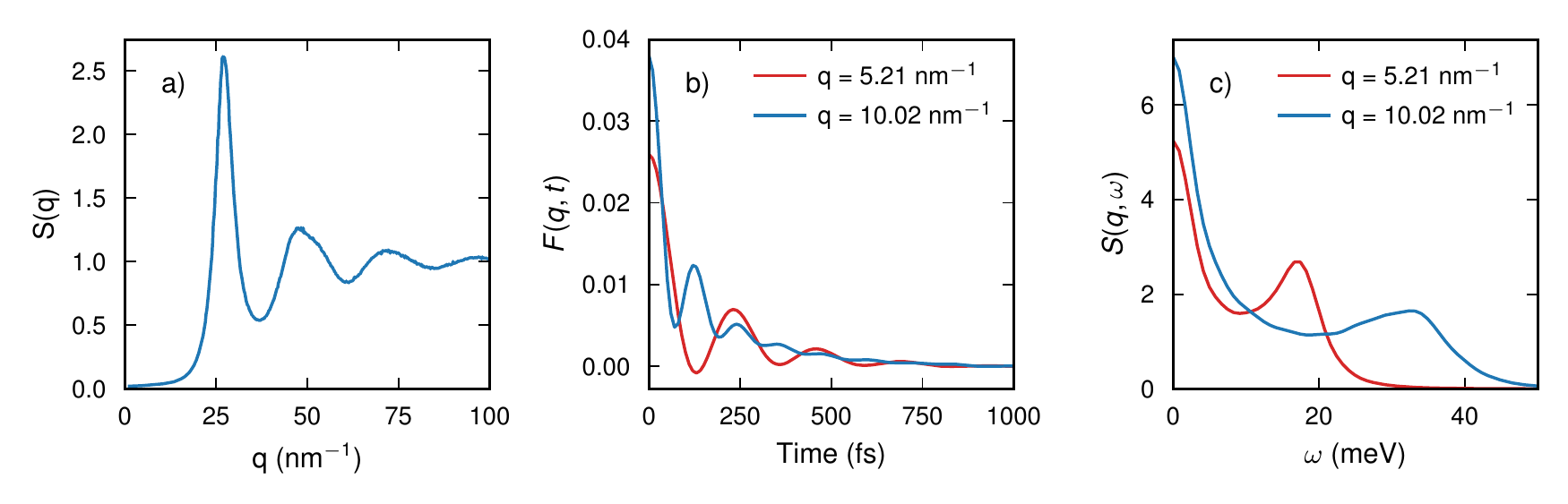}
  \caption{
  Liquid aluminum at \unit[1200]{K}.
  (a) Structure factor, (b) intermediate scattering function, $F(q, t)$, and (c) the dynamical structure factor, $S(q, \omega)$.
  }
  \label{fig:Al_liquid_panel}
\end{figure*}

To illustrate the application of \dynasor{} for analyzing liquid phases, we now consider liquid aluminum, which was simulated using the same potential as for its solid coun\-ter\-part \cite{MisFarMeh99} and using the same number of atoms (6912).
We carried out simulations at \unit[1200]{K} with isotropic $\vec{q}$-space sampling, yielding a structure factor (\autoref{fig:Al_liquid_panel}a) in good agreement with literature data \cite{MokYulKhu06}.
To illustrate the calculated intermediate scattering function, $F(q,t)$, and dynamical structure factor, $S(q,\omega)$, two slices are shown for $q=\unit[5.05]{nm}^{-1}$ and $q=\unit[9.90]{nm}^{-1}$ in \autoref{fig:Al_liquid_panel}b,c).
The behavior observed corresponds to a diffusion (gas-like) part and a vibrational (solid-like) part \cite{LinBlaGod2003}
In the time domain this corresponds to a decaying function and damped oscillator function, respectively, whereas in the frequency domain it corresponds to a decaying function that is nonzero at $\omega=0$ and a peak function, respectively.

\begin{figure*}
    \centering
    \includegraphics[width=0.95\linewidth]{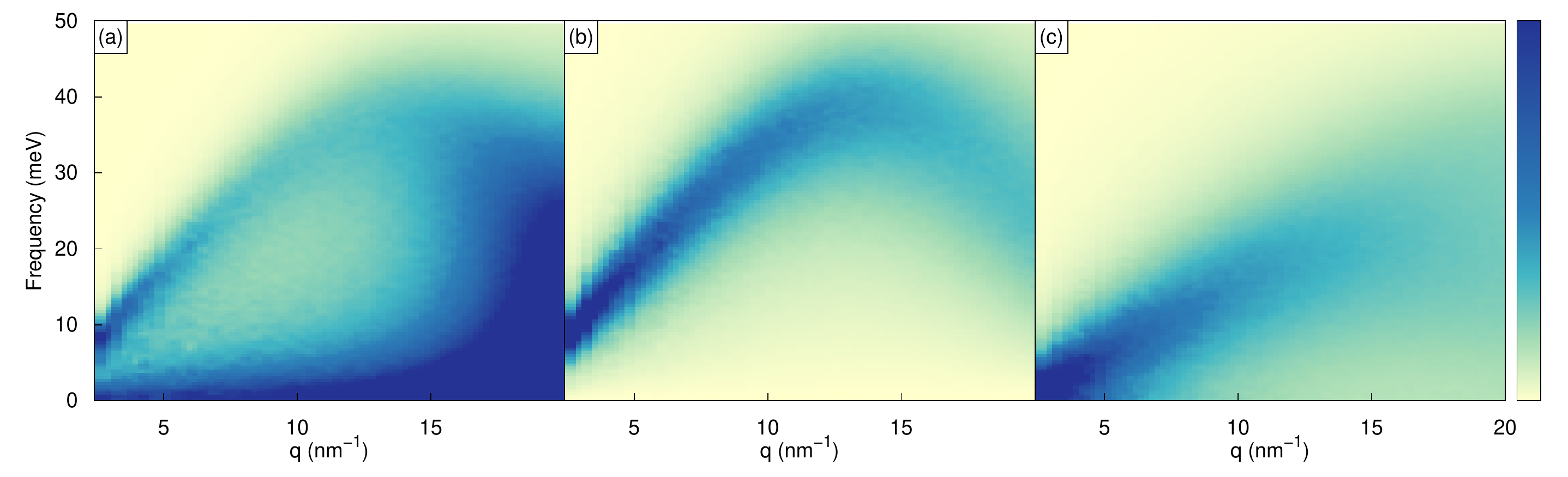}
    \caption{
        Liquid aluminum at \unit[1200]{K}.
        (a) Dynamical structure factor, (b) longitudinal current, and (c) transverse current as a function of $q$ and $\omega$.
        $\vec{q}$-vectors are cut below $\unit[2.5]{nm}^{-1}$ due to the poor resolution beyond that point.
    }
    \label{fig:Al_liquid_heatmap}
\end{figure*}

The full $q-\omega$ plane is visualized in \autoref{fig:Al_liquid_heatmap} for $S(q, \omega)$, $C_L(q, \omega)$ and $C_T(q, \omega)$.
Clear dispersion relations can be observed, which are in agreement with both experimental measurements and computer simulations \cite{MokYulKhu06}.
We note that the longitudinal dispersion, which can be seen in both $S(q, \omega)$ and $C_L(q, \omega)$, is more distinct in the latter as it does not contain the diffusive (gas-like) part.
Since resolution deteriorates for very low $q$-values, due to the finite size of the \gls{md} simulation, the x-axis has been cut at $q=\unit[2.5]{nm}^{-1}$.

\subsection{BCC titanium}
\label{sect:bcc-Ti}

Having established the basic procedure for analyzing phonon dispersion and lifetimes for a rather harmonic system such as \gls{fcc}-Al, we can now turn to a material, for which perturbative analyses fail altogether, namely the \gls{bcc} phase of titanium.
While this phase is the most stable for temperatures between 1155 and \unit[1943]{K}, the \gls{bcc} structure is dynamically unstable at zero Kelvin, leading to harmonic phonon modes with imaginary frequencies.
\Gls{bcc}-Ti therefore provides a particular interesting test case with very pronounced anharmonicity and a strongly temperature dependent phonon dispersion that has already been extensively investigated experimentally \cite{PetHeiTra91}.
The modes for which these effects are most pronounced are related to the \gls{bcc}-\gls{hcp} (TA1 at N-point) and the \gls{bcc}-$\omega$ transition (LA along H-P direction) \cite{PetHeiTra91}.
Further, \gls{bcc}-Ti exhibits spontaneous defect formation and migration \cite{FraErh20}, which complicates its analysis with lattice dynamics approaches.

The atomic interactions were described using a modified embedded-atom-method potential \cite{hennig_classical_2008}, which accurately reproduces the different phases.
\Gls{md} simulations were carried out using $12\times12\times12$ conventional \gls{bcc} unit cells.
The system was first equilibrated in isothermal-isobaric ($NPT$) ensemble in order to obtain the correct lattice parameter after which the correlation functions were sampled in the microcanonical ($NVE$) ensemble.

\begin{figure*}
    \centering
    \includegraphics{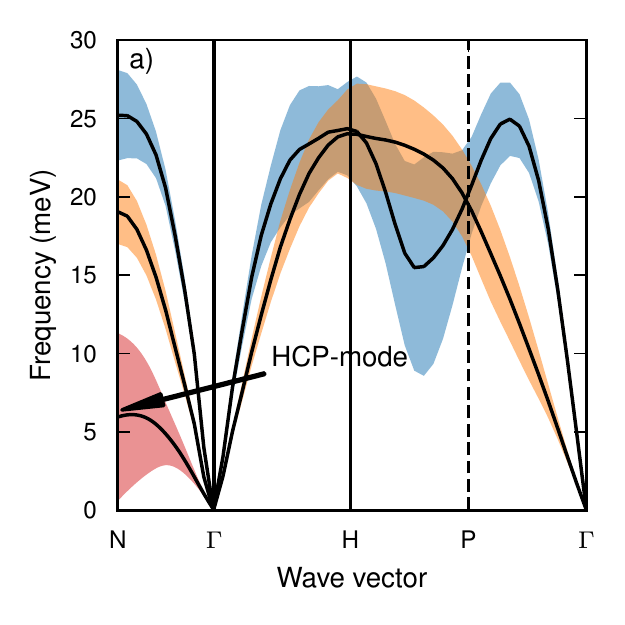}
    \includegraphics{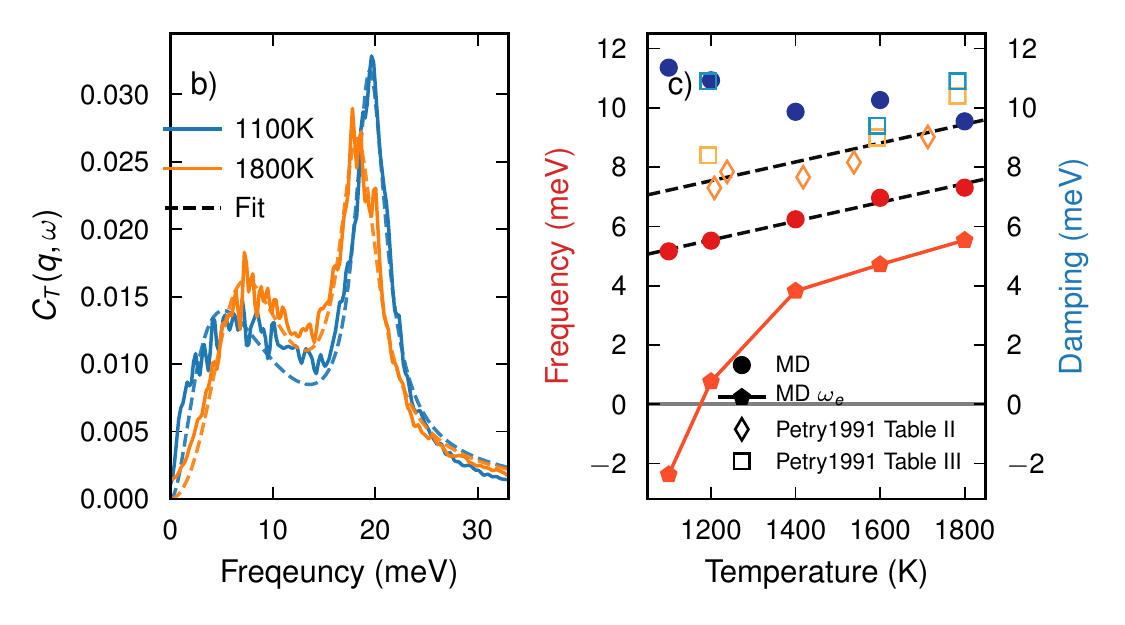}
    \caption{
        a) Phonon dispersion relation for BCC Ti at \unit[1400]{K} obtained from \gls{md} simulations.
        The shaded regions indicate the phonon lifetimes.
        b) The transverse current correlation at the N-point, and corresponding fits as dashed lines, at \unit[1100]{K} and \unit[1800]{K}.
        c) Frequency and damping of the lower most TA mode at the N-point, corresponding to the BCC$\to$HCP transition, as a function of temperature.
        The frequencies are given in red/orange and damping coefficients in blue.
        Here, results from inelastic neutron measurements \cite{PetHeiTra91} and \dynasor{} analysis are shown by open markers and filled markers, respectively.
        The frequency $\omega_e$ becomes imaginary (drawn as negative) at around \unit[1200]{K} indicating that the mode is overdamped.
        Dashed black lines are drawn as a guide  to the eye to show the linear decrease of the frequency as temperature decreases.
    }
    \label{fig:Ti_phonons}
\end{figure*}

Extraction  of the phonon dispersion and lifetimes proceeded in the same fashion as for the case of \gls{fcc}-Al.
As a result of the strong anharmonicity the correlation functions exhibit, however, much more asymmetric shapes that clearly deviate from simple Lorentzian line shapes.
The phonon dispersion at \unit[1400]{K} (in the middle of the stability range of the \gls{bcc} phase) clearly shows this strong damping, especially near the N-point and along the H-P direction (\autoref{fig:Ti_phonons}a).
The comparison with the harmonic dispersion further demonstrates the strong renormalization of the phonon modes by temperature, which not only leads to very pronounced shifts in the frequencies but also affects the shape of the dispersion, as is most apparent for the lower \gls{ta} mode along the H-P direction.

Given the importance of the TA1 mode at the N-point for the \gls{bcc}-\gls{hcp} transition \cite{PetHeiTra91}, we analyzed the temperature dependence of the transverse current correlation function at this point in more detail (\autoref{fig:Ti_phonons}b).
The results agrees very well with experimental work \cite{PetHeiTra91} both with respect to slope and absolute magnitude (\autoref{fig:Ti_phonons}c).
This illustrates how the dynamics of strongly anharmonic modes in the strongly and over-damped limits can be readily extracted using \dynasor{}.
The analysis also clarifies the distinction between $\omega_e$ and $\omega_0$ that becomes apparent for strongly damped modes.

\subsection{Liquid sodium chloride}
\label{sect:sodium-chloride}

Lastly a liquid two-component system, molten sodium chloride (NaCl), is studied in order to illustrate how the partial correlation functions can be used.
Some useful linear combinations of the partial correlation functions are the charge and mass correlation, defined as
\begin{align}
  \begin{split}
    S_\text{charge}(q,\omega) &= \sum_i \sum_j Q_i Q_j S_{ij}(q,\omega) \Big/ \sum_i\sum_j |Q_i Q_j| \\
    S_\text{mass}(q,\omega) &= \sum_i \sum_j m_i m_j S_{ij}(q,\omega) \Big/ \sum_i \sum_j m_i m_j
  \end{split},
  \label{eq:partial_combinations}
\end{align}
where $i,j$ represent the atom types, $m_i$ and $Q_i$ are respectively mass and charge of species $i$.
These linear combination are possible not only for $S(q,\omega)$ but for all correlation functions in both frequency and time domain.
Commonly acoustic type modes are revealed in mass-mass correlations whereas  charge-charge correlation can be used to investigate optical modes.

A Born-Mayer-Huggins style potential was used together with a Coulombic term as implemented in the pair style \textit{born/coul/long} in \textsc{lammps} \cite{Pli95} with the parameters reported by Lewis and Singer \cite{LewSin75}.
\gls{md} simulation were first carried out in the $NPT$ and then the $NVE$ ensemble for systems comprising 4096 atoms at \unit[1200]{K}.

\begin{figure}
    \centering
    \includegraphics[scale=0.5]{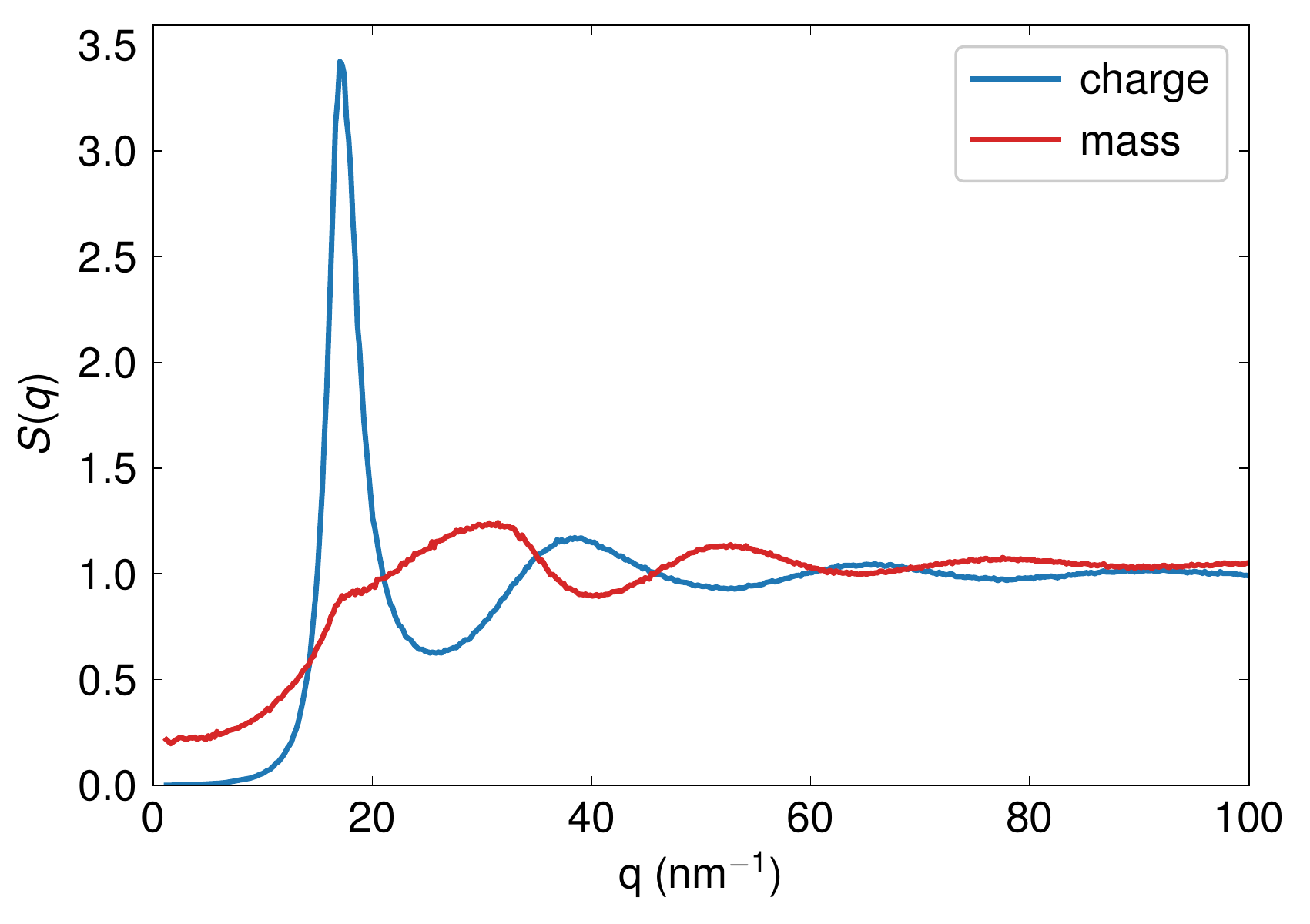}
    \caption{
        Charge and mass structure factor for liquid sodium chloride at \unit[1200]{K}.
    }
    \label{fig:NaCl_static}
\end{figure}

\begin{figure*}
    \centering
    \includegraphics[scale=0.65]{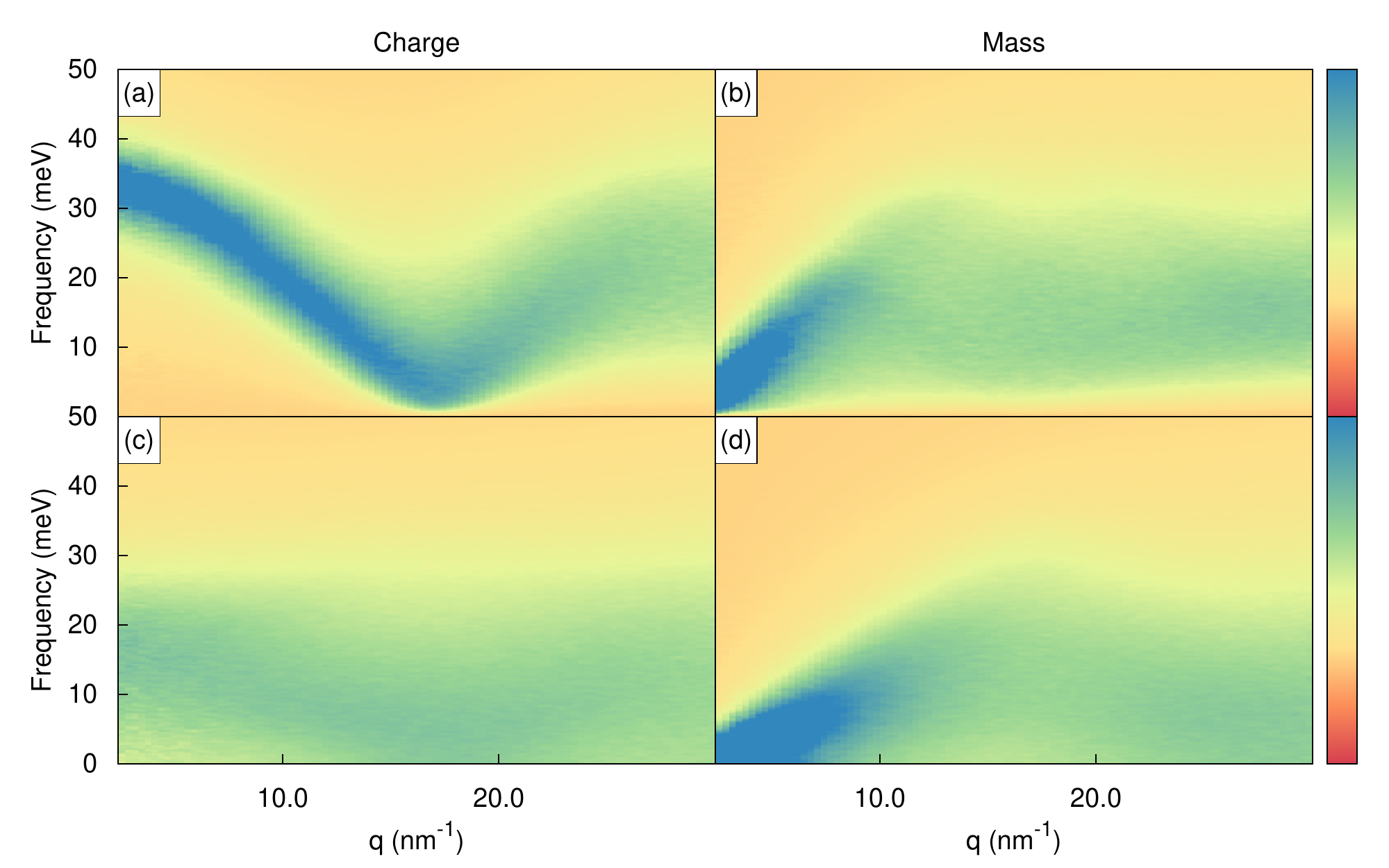}
    \caption{
        Charge (a,c) and mass (b,d) longitudinal (top) and transverse (bottom) current correlations for liquid sodium chloride at \unit[1200]{K}.
        q-points are cut below $\unit[2.5]{nm}^{-1}$  due to the poor resolution beyond that point.
    }
    \label{fig:NaCl_CMN_heatmap}
\end{figure*}

The static charge and mass structure factor is shown in \autoref{fig:NaCl_static}.
The charge and mass current correlations computed from \eqref{eq:partial_combinations} are visualized in the  $q-\omega$ space in \autoref{fig:NaCl_CMN_heatmap}.
In the charge correlation function a clear longitudinal and a weak transverse optical mode are visible, while in the mass correlation function the acoustic modes can be seen.

\section{Discussion}
All calculations in \dynasor{} are done in $\vec{q}$-space and the van Hove function $G(\vec{r},t)$ is computed as the inverse Fourier transform of $F(\vec{q},t)$ which often leads to rather poor resolution in space. The self part of both $F(\vec{q},t)$ and $S(\vec{q},\omega)$ contains information of about the self diffusion of the atoms.
However, since \gls{md} is performed studying the mean square displacement is a more efficient way of computing the diffusion constant.

\section{Conclusions}
In this paper we have presented the \dynasor{} package, which is designed to aid in the analysis of dynamical correlation functions in particular in fully or partially crystalline systems, although it is equally applicable to fully disordered systems.
We have demonstrated its usage via a few simple examples, including mono-elemental solids and liquids as well as a two component liquid.
For all these systems, current-correlation functions are shown to be very effective for analyzing the dynamics and for extracting properties such as phonon frequencies and lifetimes.

In the case \gls{fcc}-Al, for which direct comparison with standard lattice dynamical analysis techniques is possible, we demonstrate excellent agreement at low temperatures for both frequencies and lifetimes.
At higher temperatures the deviation between \gls{md} results and perturbative treatments increases.
This is a reflection of the limitations of the latter approach, which is typically terminated after the third-order \cite{Zim60, TogChaTan15} or (rarely) the fourth-order expansion term \cite{FenLinRua17, TiaSonChe18, TadTsu15}.
In contrast, \gls{md} simulations capture phonon processes to all orders.
Thereby they yield the variation of both frequencies and lifetimes with temperature without the need to resort to further approximations.

Further, we demonstrate the extraction of the temperature dependence of phonon dispersion and lifetimes for a metastable crystalline material (\gls{bcc}-Ti), for which perturbative treatments are not applicable.
Here, the results show good agreement with with inelastic neutron scattering experiments \cite{PetHeiTra91}.

The present approach of extraction dynamical correlation functions from \gls{md} simulations via \dynasor{} thus complements lattice dynamics techniques \cite{TogChaTan15, LiHellMa2014, Kon11} and is essential, e.g., when studying the vibrational properties of materials with large unit cells, low symmetry and/or strong anharmonicity, such as metastable crystals, systems with defects including surfaces and interfaces as well as amorphous and liquid systems.

\begin{acknowledgments}
This project is financially supported by the Swedish Foundation for Strategic Research (RMA 15-0062), the Swedish Research Council (2016-04342, 2018-06482) and the Knut and Alice Wallen\-berg Foundation (2014.0226).
Com\-puter time allocations by Swedish National Infrastructure for Computing at C3SE (Gothenburg), NSC (Link\"oping), and PDC (Stockholm) are gratefully acknowledged.
\end{acknowledgments}

\bibliography{lit}

\end{document}